\documentclass[letterpaper,twocolumn,10pt]{article}
\usepackage{usenix-2020-09}
\usepackage{tikz}
\usepackage{amsmath}
\usepackage[table,xcdraw]{}
\usepackage{booktabs}

\usepackage[english]{babel}
\usepackage{blindtext}
\usepackage{amsmath,amsfonts}
\usepackage{graphicx}
\usepackage{textcomp}
\usepackage{tabularx}
\usepackage{longtable}
\usepackage{array}
\usepackage{mathtools}
\usepackage{algorithm} %%%%
\usepackage{algpseudocode} %%%
\usepackage{mwe}
\usepackage{mdframed}
\usepackage{subcaption}

\newcommand\ignore[1]{}
\newcommand\ignoreA[1]{}

\newtheorem{definition}{Definition}
\usepackage{multirow}
\usepackage{enumitem}
\usepackage{listings}

\author{
{\rm Bar Meyuhas}\\
Reichman University
\and
{\rm Anat Bremler-Barr}\\
Tel-Aviv University
\and
{\rm Tal Shapira}\\
Tel-Aviv University
} % end author

\newcommand{\barm}[1]{\textcolor{blue}{Bar: #1}}

\newif\ifanswers

\begin{document}
\title{Reading Between the Strings: Breaking Down IoT Device Labeling Barriers with LLM}
\title{Beyond Strings: Advanced IoT Device Labeling with Large Language Models}
\title{Advancing IoT Device Labeling with Large Language Models: A Comprehensive Vendor and Function Identification Approach}
\title{IoT Device Labeling Using Large Language Models}

%\title{Reading Between the Strings: IoT Device Labeling Using Large Language Models}
\maketitle

\begin{abstract}
The IoT market is diverse and characterized by a multitude of vendors that support different device functions (e.g., speaker, camera, vacuum cleaner, etc.). Within this market, IoT security and observability systems use real-time identification techniques to manage these devices effectively. Most existing IoT identification solutions employ machine learning techniques that assume the IoT device, labeled by both its vendor and function, was observed during their training phase.
We tackle a key challenge in IoT labeling: how can an AI solution label an IoT device that has never been seen before and whose label is unknown? 

Our solution extracts textual features such as domain names and hostnames from network traffic, and then enriches these features using Google search data alongside catalog of vendors and device functions. The solution also integrates an auto-update mechanism that uses Large Language Models (LLMs) to update these catalogs with emerging device types. Based on the information gathered, the device's vendor is identified through string matching with the enriched features. The function is then deduced by LLMs and zero-shot classification from a predefined catalog of IoT functions.

In an evaluation of our solution on 97 unique IoT devices, our function labeling approach achieved HIT1 and HIT2 scores of 0.7 and 0.77, respectively. As far as we know, this is the first research to tackle AI-automated IoT labeling.

\ignore{ Our requirement from this labeling process is that it should not involve active requests to the devices, which in many cases are not feasible solutions.} %Instead, the labeling can be done offline, allowing the device to be observed over several hours. 

\end{abstract}

\section{Introduction}
The rapid proliferation of Internet of Things (IoT) technology in various sectors has created new security and management challenges. Correctly identifying an IoT device plays a crucial role in IoT security and observability solutions \cite{PaloAltoNetworks2023, checkpoint, firedome}. Extensive research in both industry and academia has been dedicated to IoT identification algorithms \cite{le2019policy,guo2018ip,hafeez2020iot,marchal2019audi,aneja2018iot}, which are crucial for many IoT security and observability solutions. These algorithms predominantly rely on labeled IoT datasets for the real-time identification of already \textbf{seen} devices. However, given the sheer diversity of IoT devices, collecting labeled data is a formidable challenge.

This paper introduces a method that passively monitors network traffic to automatically label unseen IoT devices by identifying their vendor and function.
To the best of our knowledge, this is the first paper to address the IoT labeling problem. The motivation to address this issue was born out of discussions with IoT security companies who are having difficulty identifying IoT devices in the wild. In the IoT realm, maintaining a lab with all known IoT devices simply isn't feasible. Moreover, the increasing number of cybersecurity attacks and new regulations make IoT vulnerability management important to various organizations nowadays. IoT security considers a vulnerability to be potential when it applies to a specific device type and vendor/model. Therefore, the first stage in any vulnerability management program is to identify and construct an inventory of assets, particularly IoT devices owned by the organization\cite{Palo_Alto_Networks}.

Labeling IoT devices with their vendor and function provides crucial visibility, ensuring administrators can effectively oversee their network. Our solution also introduces the potential for tailored security policies that can be formulated based on precise device functions, such as ones that give a device labeled 'smart doorbell' limited external communication but grant one labeled 'smart TV' broader access.

To the best of our knowledge, the high-tech industry currently relies on a single solution for IoT device labeling, with Fing \cite{fing2018devicerecognition} standing out as the leading product used by IT teams to label devices in their networks. Our solution has surpassed Fing's performance in achieving accurate results. Their approach to the labeling challenge is through crowd-sourcing. Their solution allows users, such as network owners or administrators, to label devices that the monitoring app couldn't accurately identify. However, this solution needs extensive network coverage to work optimally, and a recent study indicated its limitations in accuracy \cite{10179282}.

\ignore{
\begin{figure}[h]
\centering
\includegraphics[width=0.8\linewidth]{figures/Labeling Approach Flowchart.jpg}
\caption{IoT labeling process. The algorithm first enriched a new device data, the it identifies the vendor. The subsequent step assigns the device function, taking into consideration the obtained vendor label.}
\label{fig:labeling_diagram}
\end{figure}
}

\begin{figure*}[t]
\centering
\includegraphics[width=\linewidth]{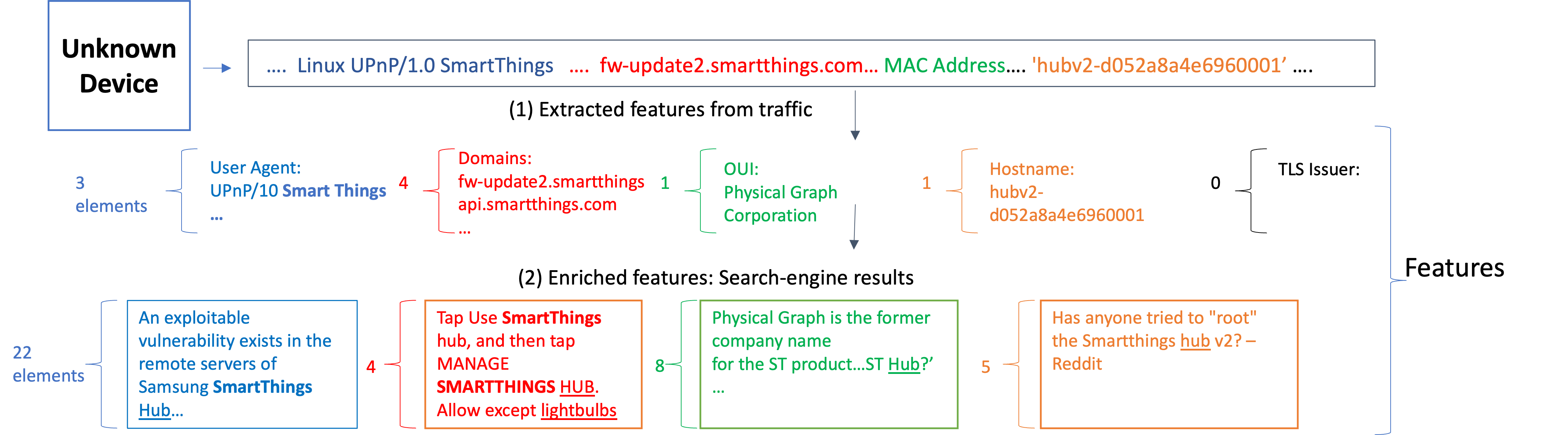}
\caption{Example of Features for the SmartThing Hub: First, we present the features derived from the traffic, followed by a sample of the enriched features (the color correlates between the feature and the enriched feature). Words relevant to the vendor label decision are highlighted in bold, and those relevant to function decisions are underlined.}
\label{samsung}
\end{figure*}

In this study, we demonstrate how recent advancements in Large Language Models (LLMs) can address the challenging task of IoT device labeling. Our algorithm uses a catalog of known vendors and the various IoT functions they produce to identify known devices. However, by harnessing the power of LLMs, it can automatically update this catalog when a new device type appears.

The algorithm starts by extracting textual strings from the network's traffic log. Specifically, it extracts the following features: domains to which the devices connect, hostnames, TLS issuers, OUI (after the MAC lookup \cite{wireshark_oui_db}), and user-agents. Note that different amounts of textual values can be associated with each feature type. For instance, a single IoT device can connect to multiple domains and may have different TLS issuers for each domain.

The algorithm then takes the text features and feeds them into a search engine so the search results' descriptions can be used to enrich the data.  Because these results typically include pages from IoT manufacturers, online shopping websites selling the device, and security blogs discussing the device's features, such searches often yield references to the vendors and functions of the IoT devices. Figure \ref{samsung} shows an example of the enriched features, and how they reveal the vendor and function. The enriched features give us a large text dataset in which we use string matching to find the vendor of the device and LLMs to determine the function of the device.  Note, however, that different enriched features can result in different labels, as seen in our example. Choosing the most accurate label among several possible ones is one of the challenges addressed by our solution. It was vital that our algorithm determine how to weigh the answers from different enriched feature values, according to the feature type. To address this challenge, we employed a machine-learning approach. 

Our labeling algorithm consists of two steps: using string matching to identify the vendor and using LLMs with zero-shot classification to label the function. In the first step, we identify the vendor by performing string matching of the text results in our dataset against a catalog of vendors and create labels based on the most common ones. In our dataset, which comprises 97 unique IoT devices from 55 distinct vendors, the algorithm achieves a high accuracy rate: 86\% for HIT1 and 89\% for HIT2~\footnote{HIT2 measures whether the correct vendor was among the top two suggested labels.}. Although there is a misconception that the OUI information can reveal the vendor, we observed that it achieves only a 64\% accuracy. This is primarily because in IoT, the OUI often identifies the NIC vendor rather than the actual device vendor.

In the second step, the algorithm receives the catalog of potential functions for the vendor identified in the first step. Here, we leverage the fact that most IoT vendors specialize in producing a limited range of device functions. We employ large language models (LLMs) and use zero-shot classification \cite{What_is_Zero-Shot} to map each feature value to its relevant functions. Specifically, we used Roberta \cite{joeddav-xlm-roberta-large-xnli} to classify each feature into possible functions and gathered the confidence scores of these classifications. Our method then aggregates these scores to establish an overall confidence level for each label. The label with the highest confidence level is then assigned as the device's function. In our dataset, which includes 21 functions, the algorithm achieves an accuracy of 70\% for HIT1 and 77\% for HIT2. Since our algorithm works on textual features, it can provide explanations to humans who wish to verify its result, by outputting the set of features upon which its decision was based. 
     \ifanswers

 The rest of the paper is organized as follows:
\begin{itemize}
     \item Section \ref{sec:requirments} outlines our IoT labeling method requirements, followed by a technical background in Section \ref{sec:background}.
     \item The catalogs used for labeling are introduced in Section \ref{sec:list}, with the catalog updating mechanism discussed in Section \ref{sec:catalog_acquistion}. 
     \item In Section \ref{sec:dataset} and \ref{sec:features_and_enrichment} we present our dataset and the enriched features.
     \item Section \ref{sec:methods} and \ref{sec:experiments} presents the labeling algorithm and our experiments results.
 \end{itemize}
         \fi

\ifanswers

\begin{figure}[ht]
\centering

\includegraphics[width=0.95\linewidth]{figures/Arch System Diagram1.png}
\caption{A schematic illustration of a system integrating our IoT labeling solution (highlighted in purple). It elucidates the process from online device identification to offline label extraction and IoT Identification training upon detection of a new device.}
\label{fig:architecture}
\end{figure}
\fi
\section{IoT Labeling  Requirements  }
\label{sec:requirments}
The requirements we outlined for an efficient and automated IoT labeling solution are as follows:
\begin{itemize}
    \item \textbf{Universality:} The algorithm is aimed at processing data from IoT devices it has never before encountered. 
   The algorithm leverages the use of a catalog that contains possible vendors and functions they support. The algorithm is able to update this catalog when the device type does not already appear in the catalog. 
   \item \textbf{Accuracy:} High accuracy is paramount in IoT labeling due to its critical role in enhancing network observability and security.
\item \textbf{Explainability:} The solution incorporates a confidence level and justification for each label, allowing for potential human verification. This aspect caters to scenarios where it is better to have an inaccurate label than one that is incorrect.
\item \textbf{Passive solution:} 
The IoT labeling operates without actively probing or querying the device for information, as is done in Shodan \cite{shodan_website} and Censys \cite{censys_website}. We note that in many cases, this is not a possible or informative solution. 

\item \textbf{Offline process:} The IoT labeling operates offline by collecting the features over long periods (hours or days). The algorithm labeling running time is not a critical parameter since it is run only when it encounters a new device that was not recognized by the system. 

\end{itemize}

\ifanswers
The IoT labeling highlighted in Figure \ref{fig:architecture} is an offline task, as opposed to the IoT identification inference, which is an online task. In a common IoT security architecture, the online system should be able to identify known devices using one of the existing IoT identification algorithms  \cite{le2019policy,guo2018ip,hafeez2020iot,marchal2019audi,aneja2018iot}. 
If the device cannot be identified, and the device is IoT \footnote{There are algorithms to classify the device as an IoT type as opposed to general purpose computer \cite{9110451}}, then our IoT labeling algorithm used to determine the label. In this case, the identification algorithm trains again with the new data from the IoT and its label, and the identification inference algorithm is updated.  
\fi

\begin{figure}[h]
\centering
\includegraphics[width=0.92\linewidth]{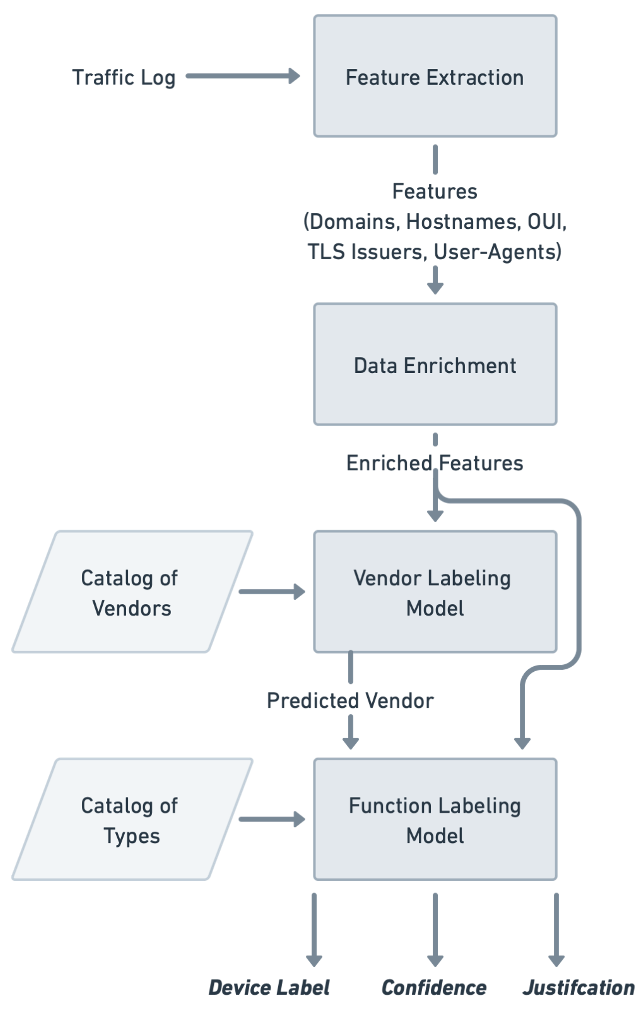}
\caption{A schematic illustration of our IoT labeling solution. First, features are being extracted and then enriched. Second, we perform our vendor and function models labeling. The system's output is label, confidence and justification for each device.}
\label{fig:labeling_solution}
\end{figure}

\section{Background: Zero-Shot Classification}
\label{sec:background}

In traditional ML classification tasks, a model is trained on a specific set of classes using samples from each type. In our work, each of these classes is equivalent to a new IoT type.  If a new class emerges after the model has been trained, the entire model either needs to be retrained from scratch or fine-tuned, necessitating new labeled data for that class. This iterative process is both time consuming and labor intensive, which makes it not sustainable, especially in dynamic environments like IoT. In the IoT realm, it is generally not feasible to obtain samples from every type, which means the set of classes will evolve and expand over time.

Zero-shot classification is unique in that it classifies text without requiring explicit training on the target classes. Instead of learning representations only for classes that were seen during the training phase, zero-shot classification leverages the semantic relationships between classes. It often uses auxiliary information such as class attributes or textual descriptions to perform the classification. This enables the model to infer classes it has never seen during training, create new labels, and efficiently update the model with new classes. In our case, the IoT catalogs that serve as the input to our algorithm need to be updated based on the zero-shot classification; however, the model itself and its optimized parameters do not require re-training once the catalog has been updated.

\ignore{
Our decision to utilize zero-shot classification was supported by its inherent ability to support varied class selections,that way, we could choose as classes, different sets of device functions. A prime example of this in the IoT domain, where knowing the vendor of a device might influence the set of potential functions that the device could have. That way, we could use the zero-shot model with no post-change training. It also allows us to add more labels as required when if a new device function will be manufactured. Furthermore, a pervasive challenge in IoT classification is the lack of labeled data given the sheer diversity of IoT devices.}
In essence, zero-shot classification, with its ability to dynamically adapt to new classes and its minimal reliance on exhaustive labeled data, provides a solution to the challenges of IoT labeling.

\section{Dataset}
\label{sec:dataset}
We trained and tested our IoT labeling algorithm on a  dataset that amalgamates five distinct open-source datasets (\cite{sentinel_dataset},\cite{dataset_imc}, \cite{sivanthan_dataset}, \cite{yourthings_dataset}, and  \cite{ourlab_dataset}).  These datasets were selected by previous research to provide a representation for identification-oriented tasks.   The dataset includes traffic logs of 161 IoT devices~\footnote{Each device can be identified uniquely by its MAC address}, from 55 distinct vendors, categorized across 21 different functions. We note that the traffic logs are of different lengths, varying from hours to days.  

For our testing, we randomly selected a single device from each group of devices that shared the same type (i.e., vendor and function), in order to avoid bias arising from multiple repeated instances. All the subsequent experiments discussed in this paper were based on analyses conducted on this collection of 97 unique devices that appear in the dataset, unless explicitly stated otherwise.
Nonetheless, our experiments showed that devices of the same type often have lower similarity when it comes to feature values, as detailed in Section \ref{sec:data_collection_non_reptative}. This outcome likely occurred because a single device type can encompass a range of different IoT models. For example, you can have various camera models from the same IoT vendor. Because we lacked the information to determine whether the IoT devices shared the same model, we decided to take the restrictive approach and include only one device per type.  

\ignore{ As far as our research extends, this collection constitutes the broadest variety of devices cited in academic literature for such purposes.
}

The devices in our datasets are labeled with vendor and function since they were recorded in the lab. An example of a label with vendor and function would be: Belkin Plug. By manually examining the device's name on the internet, we were able to translate it into a function.  This labeled dataset provided us with a ground truth basis for our subsequent experiments and models.

\subsection{ Enriched Features}

\ignore{ 
\begin{figure*}
     \centering
     \begin{subfigure}[t]{0.45\linewidth}
         \centering
         \includegraphics[width=\linewidth]{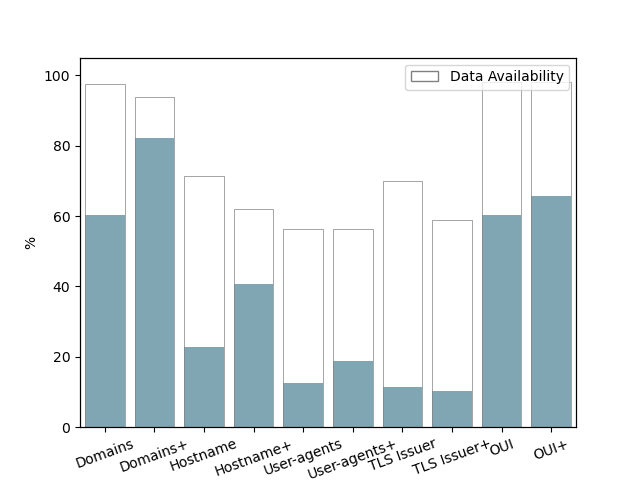}
         \caption{Vendor Labeling with  String-matching Algorithm.}
        \label{fig:stringmatching_vendor_results}
     \end{subfigure}
     \begin{subfigure}[t]{0.45\linewidth}
         \centering
         \includegraphics[width=\linewidth]{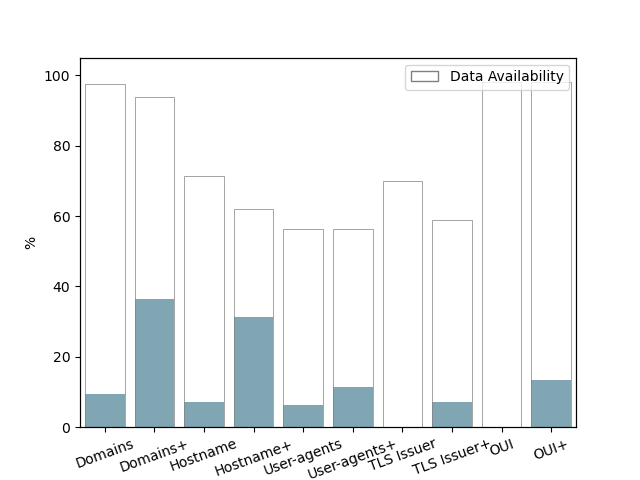}
         \caption{Function Labeling with Roberta Algorithm.}
         \label{fig:nlp_type_results}
     \end{subfigure}
     \hfill
     
     \caption{
Comparative analyses of the availability and accuracy of features for predicting the function (\ref{fig:nlp_type_results}) and the vendor (\ref{fig:stringmatching_vendor_results}) of IoT devices. }
\label{fig:availability_and_accuacy_features}
\end{figure*}
  }

\begin{figure}[ht]
    \centering
    \includegraphics[width=\linewidth]{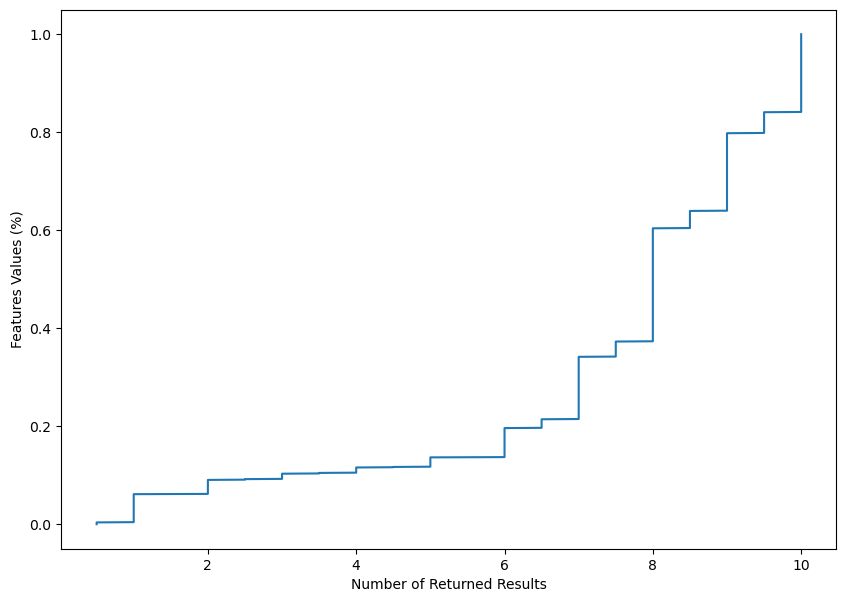}
    \caption{CDF distribution showing the number of results returned per feature value}
    \label{fig:cdf_query_distribution}
\end{figure}

Our labeling algorithm treats and processes the traffic from each IoT device separately. Based on the traffic, it derives all the feature values from all the feature types (hostname, domains, TLS Issuers, OUI, User-Agents). The feature values then undergo an enrichment process in which a search query is performed for each feature value; our solution extracts at most the top 10 results from this query. We executed our data enrichment phase by using SerpAPI \cite{serpapi} to access Google search results across the web.  During this process, most of the feature values returned less than 10 results, but 60\% returned more than 8 results, as shown in Figure \ref{fig:cdf_query_distribution}.

The structure of each search result from SerpAPI \cite{serpapi} is composed of a title and snippet of the web page; both of these were used as enriched data. Google uses a number of different sources to automatically determine the title of the search result. The snippets are automatically created from the page content\cite{google_dev}, which usually contains the search query itself.
In our case, the search result data often encompassed sites owned by the device vendor and occasionally linked to public forums where device users discuss their device's network behavior. In some cases, we even found links to eCommerce platforms that sell devices, which enabled the identification of specific device models. 

We denote the enriched version of feature type $f$, by $f^{+}$.  Correspondingly, we define the vector of the enriched feature as follows:

\begin{definition}
    Let $f^+_{t,p}$ be the vector $= ( f^+_{t,p,1} \dots f^+_{t,p,k} )$ of the enriched feature values vector with $k$ top search results on feature value $f_{t,p}$ of feature type $t$. $f^+_{t,m,j}$ is the $j$-th enriched feature returned from a search on  $f_{t,m}$ feature $t$ value.
\end{definition}
 
Table \ref{tab:term} summarizes the terminology used throughout the paper. All the devices in the dataset, including all the features with their enrichment data, are publicly available in JSON format at \cite{our_dataset_JSON} for further research into IoT device labeling. The format of the JSON is given in Appendix, Table \ref{tab:term}.

% ANAT HOW YOU RECEIVE LIST
% FIGURE ABOUT THE LIST
\section{Labeling Algorithm}
\label{sec:methods}

 \begin{table*}[ht]
\centering
\begin{tabular}{|c|c|}
\hline
\textbf{Term} & \textbf{Description} \\
\hline
\hline
$V = \{V_1,V_2,...V_{n_V}\}$ & Catalog of Vendors of length $n_V$ \\
\hline
$F = \{F_1,V_2,...F_{n_F}\}$ & Catalog of functions of length $n_F$ \\
\hline
$T = \{T_1,T_2,...T_{n_T}\}$ & Catalog of types of length $n_F$  where $T \subseteq V \times F$ \\
\hline
\hline
$FT=\{hostname,domains,TLS,OUI, UserAgent\}$ & Feature type set (size of 5)  \\
\hline
 $f_{t} = (f_{t,1}...f_{t,m})$ &  Feature values vector of feature type $t\in TF$\\
\hline
$f^+_{t,p} = ( f^+_{t,p,1} \dots f^+_{t,p,k} )$ & Enriched feature values
 vector with $k$ top search result on $f_{t,p}$\\
\hline
\hline
$S_{F_m,f^+_{t,p,l}}$  & Confidence scores labeling $f^+_{t,p,l}$  with $F_m$ \\
\ignore{\hline
$\theta $ & Confidence threshold   \\}
\hline
 $w_t$ & Weight  per feature type $t\in FT$ \\
\hline
$ \theta_t$
 & Confidence threshold  per feature type $t\in \theta$ \\
\hline
\end{tabular}
\caption{Terms and Terminology for Labeling Algorithms and Scoring Techniques}
\label{tab:term}
\end{table*}

In this section, we present the labeling algorithm. To determine the device type, we first determine the vendor label and then find the function by examining which IoT device functions the identified vendor manufactures. Vendor and function labeling essentially operate in similar ways. Each enriched feature for a single IoT device is matched against the corresponding vendor or function catalog ~\footnote{We focus only on enriched features since our experiments show that adding information from the basic features does not improve accuracy, mainly because the search value typically appears in the enriched data.}. We checked several matching algorithms in our experiments, including string matching, Roberta model \cite{conneau2019unsupervised}, and ChatGPT \cite{OPEN_AI_gpt}.
When it comes to vendor labeling, the naive string matching provided the best performance, while the Roberta model outperformed the other options for  function labeling. Although Roberta is adept at handling context-driven classification, it's not ideal for vendor labeling. This is primarily due to two reasons. First, zero-shot classification is not designed to effectively handle hundreds of distinct labels, as in the case of vendors. Second, the vendor classification task depends less on context and more on specific strings, making string-matching approaches more suitable. 
The Roberta model is suitable to use context and semantic similarity for the function labels because (a) it is based on Transformer architecture \cite{joeddav-xlm-roberta-large-xnli} and (b) the function catalog size is small enough for it to handle.

Next, for each enriched feature value (i.e., search result) and potential label, the algorithm produces a confidence score indicating the likelihood that the enriched feature value correctly classified the potential label.

Formally:

\begin{definition}
Let \( S_{label,f^+_{t,m,j}} \) denote the confidence score that the enriched feature value \( f^+_{t,i,j} \) corresponds to the \( label \). The \( label \) can be a vendor label, i.e., \( V_m \in V \) in vendor labeling or can be a function label, i.e., \( F_m \in F \) in function labeling.
\end{definition}

In vendor labeling, the algorithm produces a confidence score based on the number of times a label correctly matches the enriched feature. 

For the function labeling, we used the Roberta confidence score \cite{conneau2019unsupervised}. To enhance accuracy, the algorithm performs the labels of only functions associated with the identified vendor, when available; otherwise, it performs full check with all the functions in the catalog.

Next, our algorithm calculates an aggregated score for each potential label. This is done by weighting the confidence score based on the original feature type of the enriched feature and the number of results returned for that feature type. The optimization of this scoring is detailed in the next section. The label with the highest score is then selected.

\begin{algorithm}
\caption{Vendor Labeling Algorithm}
\begin{algorithmic}[1]
\State \textbf{Input}: 
A traffic log of an IoT device with enriched feature values $f^+_{t,i,j}$ for all possible value types $t \in FT$.
\State \textbf{Perform}: 
\For{each label vendor $V_m  \in V$}
    \For{every enriched feature value $f^+_{t,i,j}$}
        \State Let $S_{V_m,f^+_{t,m,j}}$ be the confidence score, 
        \State the number of times the string appears \State  $V_m$ at the enriched feature $f^+_{t,i,j}$.
    \EndFor
   \State Calculate the aggregate score of $V_m$ based on the confidence score across all enriched feature values and types.
\EndFor
\State \Return the vendor label with the highest score.
\end{algorithmic}
\label{alg:Vendorlabeling}
\end{algorithm}

\begin{algorithm}[ht]
\caption{Algorithm for Function labeling}
\label{alg:functionlabeling}

\begin{algorithmic}[1]
\State \textbf{Input}: 
\State A traffic log of an IoT device with enriched feature values $f^+_{t,i,j}$ for all possible value types $t \in FT$.
\State Let $Vendor$ be the identified vendor after the vendor labeling.

\State \textbf{Perform}: 
\State Let \( FL \) be a set of candidate relevant function labels in
\State type catalog \( FL = \{ F_m \mid (vendor, F_m) \in T \} \).
\State If $FL = \emptyset$ then $FL = F$, i.e., all possible functions.
\State Calculate the confidence values:
 \For{each label function $F_m  \in FL$}
   \For{each enriched feature value \( f^+_{t,i,j} \)}
     \State Given the Roberta model to labeling \( f^+_{t,m,j} \)
     \State against possible labels $FL$.
     \State Let \( S_{F_m,f^+_{t,m,j}} \) be the confidence score of \( F_m \in FL \).
   \EndFor
   \State Aggregate the score of $F_m$ based on the confidence score across all enriched feature values and types.
 \EndFor

 \State \Return $Function$, the function label with the highest score.
 \State \Return the type of the IoT,  (Vendor,Function).
\end{algorithmic}
\end{algorithm}

To assess the contribution of each feature type to the overall accuracy, we calculated the accuracy, which is the ratio of correct device labels obtained when running the labeling algorithm on each specific feature type.
 Figure \ref{fig:stringmatching_vendor_results} shows the accuracy of string matching in predicting the vendor and Figure \ref{fig:nlp_type_results}
 shows the accuracy of Roberta in predicting the device function.  As can be observed, different enriched features exhibit varying levels of accuracy, where Domains$^{+}$ and Hostname$^{+}$ have the highest accuracy in both vendor and function labeling algorithms. Moreover, as we noted before, the enriched feature is more informative than the basic feature. We also found that it is harder to predict the function than the vendor. One clear outcome is that we must weight the label differently according to the feature type and concentrate on the enriched features.

Our vendor labeling algorithm is outlined in Algorithm \ref{alg:Vendorlabeling}, and our function labeling algorithm is detailed in Algorithm \ref{alg:functionlabeling}.  

\begin{figure*}[h]
     \centering
     \begin{subfigure}[t]{0.45\linewidth}
         \centering
         \includegraphics[width=\linewidth]{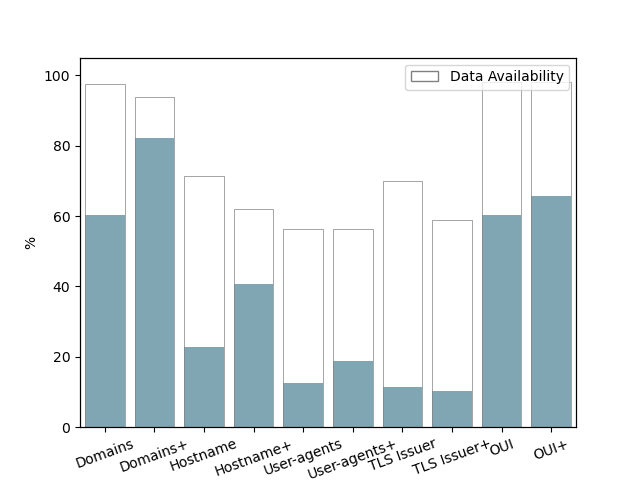}
         \caption{Vendor Labeling with  String-matching Algorithm.}
        \label{fig:stringmatching_vendor_results}
     \end{subfigure}
     \begin{subfigure}[t]{0.45\linewidth}
         \centering
         \includegraphics[width=\linewidth]{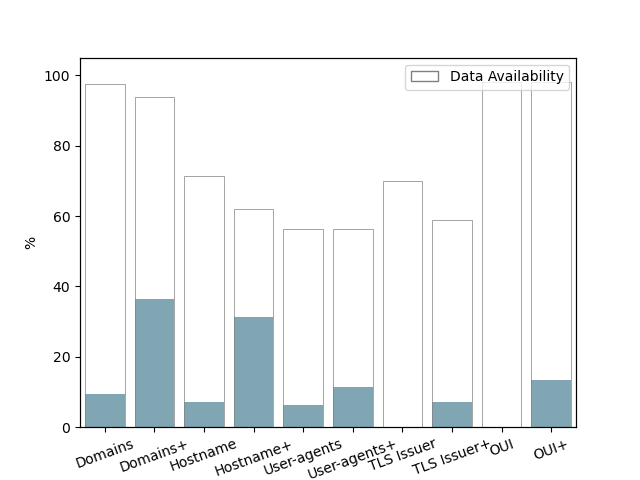}
         \caption{Function Labeling with Roberta Algorithm.}
         \label{fig:nlp_type_results}
     \end{subfigure}
     \hfill
     
     \caption{
Comparative analyses of labeling accuracy per feature, indicated by filled bars and availability, indicated by hollow bars (represents the percentage that the feature exists across the dataset). Figure \ref{fig:nlp_type_results} presents the accuracy of the device function while Figure \ref{fig:availability_and_accuacy_features} presents the vendor. }
\label{fig:availability_and_accuacy_features}
\end{figure*}
\section{Experiments Results}
\label{sec:experiments}
\begin{table*}
\centering
\resizebox{2\columnwidth}{!}{%
\begin{tabular}{|l|c|l||c|c|l|}
\hline
Method & Catalog of Vendors & Features & Accuracy  HIT1 & Accuracy HIT2 \\%& Empty Result \\
\hline
\hline
OUI & N & MAC & 0.64 & - \\%& - \\
\hline
\textbf{Our IoT labeling} & Y & \textbf{Domains$^{+}$,Hostname$^{+}$,TLS$^{+}$,User-Agents$^{+}$},\textbf{OUI$^{+}$} & \textbf{\begin{tabular}[x]{@{}c@{}} 0.86\\(0.89) \end{tabular}} & \textbf{\begin{tabular}[x]{@{}c@{}} 0.89\\(0.92) \end{tabular}} \\%& 0.006 \\
\hline
Our IoT labeling & Acquired & {Domains$^{+}$,Hostname$^{+}$,TLS$^{+}$,User-Agents$^{+}$},{OUI$^{+}$} & 0.76 & 0.8 \\%& 0.006 \\ 
\hline
GPT-4 & N & Domains, Hostname, TLS, User-Agents, OUI & 0.83 & 0.86 \\%& 0.006 \\
\hline
Fing & N & Hostname, User-Agents, MAC & 0.76 & - \\
\hline
\end{tabular}%
}
\caption{Accuracy of vendor labeling, where the catalog of vendors is given (Y-yes) or (N-not) or Acquired on-the-fly. The accuracy results for the whole dataset appear in parentheses}
\label{tab:vendor-experiments}
\end{table*}

In this section, we present the experimental results of our IoT labeling.  
We compare our solution to a few other common methods that are common in industry and academia. First is Fing product \cite{fing2018devicerecognition}, a labeling solution that is common in the industry, and is used by leading cyber-security firms. Second is the OUI method, based on the MAC address. In addition, we present a few versions of our method with more naive approaches. 
To evaluate the effectiveness of our methods, we present HIT1 which is the ratio of devices for which the method correctly predicted its label to the total number of devices in the dataset. Similarly, we present HIT2, the ratio of devices for which the correct label is in the top two labels predicted.
We also present the ratio of devices that could not be labeled, where the algorithm did not return any results, for example, if there were no features or no enriched features above the threshold. In all the techniques that required optimization learning, we used 5-fold cross-validation to ensure the training and testing were done on different devices. 

We found a mostly naive approach to vendor labeling that uses OUI, the first 24 bits of a MAC address that identify the vendor of a network interface card (NIC) or network adapter \cite{wireshark_oui_db}. This technique does not require the vendor catalog but has a lower accuracy of 0.64 since it often identifies the NIC vendor, which is usually different from IoT vendor.
We also checked GPT-4's ability to predict the vendor based on device features. Given the limited number of tokens that can be used with this large language model, we restricted our input to the original dataset features, without enrichment.  Because GPT-4  learns the Internet, it has enrichment information at some level.
\ifanswers
The prompt\footnote{Described in detail in Appendix \ref{app:prompt}} instructs the model to label the IoT devices' vendors, assign a confidence score to each label, and provide concise justifications (explainability) limited to 50 words, for its classifications.
\fi

\begin{figure}
    \centering
    \includegraphics[width=\linewidth]{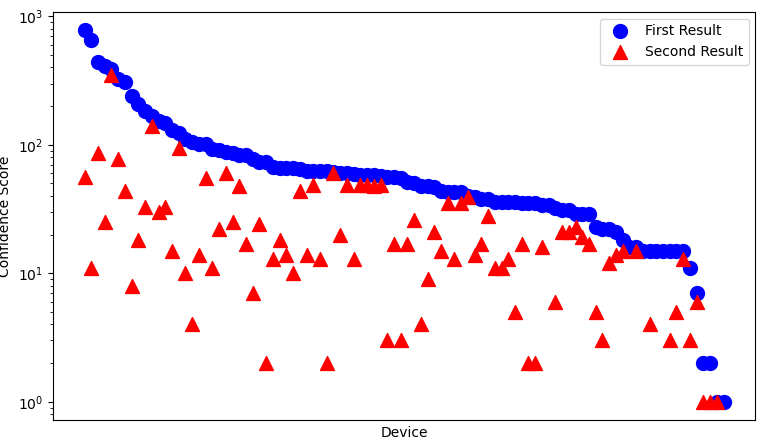}
    \caption{Confidence score per device of our vendor labeling algorithm (equals to the number of label matching in the enriched data), ordered by the first score matching. The highest score is marked in a circle and the second highest is marked in a triangle.  }
    \label{fig:enter-label}
\end{figure}

Our vendor labeling uses string matching based on all the enriched features. This method yielded the highest results with an HIT1 accuracy of 0.86 and an HIT2 accuracy of 0.89.
The GPT-4 model, without the assistance of a vendor catalog, achieved a respectable accuracy of 0.83 (HIT1) and 0.86 (HIT2). This suggests that while GPT-4's contextual understanding is excellent, more straightforward methods like string matching outperform it for vendor labeling. 

Figure \ref{fig:enter-label} shows the confidence score in our vendor labeling. It is evident that these are definite results, with a high occurrence of the string in the data; in most cases, the highest score is much higher than the second score. Note that the y-axis is in log scale.

\begin{figure}
    \centering
    \includegraphics[width=\linewidth]{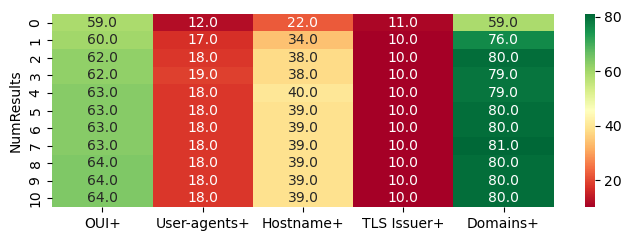}
    \caption{The distribution of correct results of vendor labeling for various features using the string matching method. Darker shades indicate higher numbers of correct results. }
    \label{fig:string_matching_across_features_num_results}
\end{figure}

We conducted an experiment to measure the accuracy of our labeling algorithm across each feature and gauge the impact of different features on the method's success. In Figure \ref{fig:string_matching_across_features_num_results} we show the number of correct results as a function of the number of returned results, where zero indicates no enrichment. It is evident that while enrichment enhances the labeling of vendors using the string matching method, the  improvement is not as substantial compared to how much enrichment impacted the function labeling method. Furthermore, some features, such as TLS Issuer, appear to be less affected, and enrichment does not elevate their performance at all.  We did not see any notable improvement for vendor labeling, when we performed the same optimization (weighting features), described earlier this section.

\begin{table*}[ht]
\centering
\resizebox{2\columnwidth}{!}{%
\begin{tabular}{|l|c|l||c|c|l|}
\hline
  Method  &  Catalog of Functions &  Features &  Accuracy HIT1 & Accuracy HIT2 \\ %& Empty Result\\
\hline
\hline
 String Matching  & A & Domains$^{+}$, Hostname$^{+}$ , TLS$^{+}$, User-Agents$^{+}$, OUI$^{+}$& 0.49 & 0.63 \\%& 0.006 \\
\hline
Roberta  &  A & Domains$^{+}$, Hostname$^{+}$, TLS$^{+}$ , User-Agents$^{+}$, OUI$^{+}$ & 0.56 & 0.65 \\%& 0.006 \\
\hline 
Roberta  &  A & Domains, Hostname, TLS, User-Agents, OUI& 0.16 & 0.2 \\%& 0.006 \\
\hline 
GPT-4 & A & Domains, Hostname, TLS, User-Agents, OUI & 0.65  & 0.75 \\%& 0.006 \\
 \hline
 GPT-4 & V & Domains, Hostname, TLS, User-Agents, OUI & 0.67  & 0.81 \\%& 0.006 \\
 \hline
 %Our Function Labeling &  V & Domains$^{+}$, Hostname$^{+}$, TLS$^{+}$ ,User-Agents$^{+}$, OUI$^{+}$& {0.69} (0.75) & {0.85} (0.875) & 0.06 \\
Fing & N & Hostname, User-Agents, MAC & 0.57  & - \\
\hline
\textbf{Our Function Labeling }  &   V & \textbf{Domains}$^{+}$, \textbf{Hostname}$^{+}$, \textbf{TLS}$^{+}$ , \textbf{User-Agents}$^{+}$, \textbf{OUI$^{+}$} & \textbf{\begin{tabular}[x]{@{}c@{}} 0.7\\(0.74) \end{tabular}}  & \textbf{\begin{tabular}[x]{@{}c@{}} 0.77\\(0.84) \end{tabular}}\\% & 0.02 \\
\hline
 %\textbf{GPT-4} & V&\textbf{Domains, Hostname, TLS }& \textbf{0.79} & \textbf{0.84} & 0\\
 %\hline
%\hline
\end{tabular}%
}
\caption{Function Labeling Accuracy. 'A' denotes accuracy when all items in the function catalog are given, while 'V' represents accuracy when only the functions associated with the device vendor are provided.}
\label{tab:function-experiments}
\end{table*} 

%\subsection{Function Labeling Comparison}
Table \ref{tab:function-experiments} shows the results for function labeling. We compared our method's results with several variations of labeling method using more simple string matching or advanced (GPT-4) methods.
When analyzing the performance of the methods using the full catalog of functions, Roberta performs much better than the string matching method. We also observed that without enriched features, Roberta performance is very low (HIT1 accuracy of 0.16). However, with enriched features, Roberta exhibits a commendable improvement, achieving an HIT1 accuracy of 0.56 and an HIT2 accuracy of 0.65.
When functions are restricted to those associated with the vendor (represented by a 'V' in Table \ref{tab:function-experiments}), the algorithm receives the catalog of potential functions for the identified vendor, and leverages the fact that most IoT vendors specialize in producing a limited range of functions. The GPT-4 model, without the assistance of a type catalog, achieved an accuracy of 0.65 (HIT1). When we used the GPT model together with the type catalog, there was a slight improvement. We assume that the type catalog had only a slight impact on the GPT model because GPT had this knowledge when labeling, even without the type catalog.  %\anat{CAN WE CHECK CHATGPT WITH CATALOG} 
Roberta, based on the types catalog using all enriched features and weighted using our optimization process, surfaces as the top performer with an HIT1 accuracy of 0.7 and an HIT2 accuracy of 0.77; this significantly outperformed the string matching method.

\textbf{Accuracy of Type Labeling}
The accuracy of type labeling refers to the successful labeling of both a device's vendor and function. Utilizing a combination of string matching for vednor labeling and Roberta for function labeling, the best accurate method for each task, yield a HIT1 accuracy of 0.6, a HIT2 accuracy of 0.7.

%In Table \ref{tab:type-experiments}, we present the results of labeling using string matching for the vendor and Roberta for the function.

\ignore{
\begin{table}[ht]
\centering
\caption{Type Labeling Accuracy. Our labeling method is composed of vendor and function labeling, we employed for it two distinct methods: string matching and Roberta.}
\resizebox{\columnwidth}{!}{%
\begin{tabular}{|c||c|c|c|}
\hline
Method & Accuracy \newline HIT1 & Accuracy \newline HIT2 & Empty \newline Result \\
\hline
\hline
\parbox[t]{2cm}{String-Matching \\ \& Roberta} & \textbf{0.6} & \textbf{0.7} & 0.05 \\
%\hline
%Best String matching for vendor with Best GPT-4 for function & ? & ? & ? \\
\hline
%GPT-4  & 0.71 & 0.74 & 0.04 \\
%\hline
\end{tabular}
}
\label{tab:type-experiments}
\end{table}}

\textbf{Results for All vs. Unique Devices}
\label{sec:data_collection_non_reptative}
As mentioned in Section \ref{sec:dataset}, we randomly selected devices for type in order to avoid repetitive devices that may bias our results. In this subsection, we show the results of our method on all the devices alongside a description of the diversity of the data  within devices of the same vendor and function. 
In Tables \ref{tab:vendor-experiments} and \ref{tab:function-experiments} we present the results of our labeling method on \textit{all} the devices in parentheses. Our function labeling method achieved 0.74 for HIT1 and 0.85 for HIT2. While our vendor labeling resulted in 0.89 and 0.92 for HIT1 and HIT2, respectively.

To verify that no recurring devices were included in our comprehensive collection, we employed the Jaccard similarity coefficient, which quantifies the degree of overlap between two sets. Each device was scrutinized for similarities across all features. This process allowed us to calculate an average similarity for each pair of devices based on these keys. This process facilitated the calculation of a similarity measure within groups of devices sharing the same vendor and function. Devices from the same vendor and function are naturally expected to share certain similarities, yet they can still exhibit distinguishing patterns due to different factors such as device model, hardware configuration, and software version.
%Furthermore, we computed the standard deviation to provide insights into the dispersion of these similarities.

Our findings, as depicted in Table \ref{tab:similarity}, confirmed that the average similarity between devices was reasonably low, thereby suggesting a high degree of uniqueness across the samples in the same group.
%We also calculated the cumulative distribution function (CDF) of hostname similarities, which revealed that approximately 78\% of the compared hostnames are distinct.

%\barm{move it to the experiment section, add an explanation on an experiment without the repetitive}
\begin{table}
\centering
\caption{Similarity Measures for Features. This table summarizes the average similarity and standard variation for features within groups of devices (same vendor and function).}
%\resizebox{\columnwidth}{!}{%
\begin{tabular}{lc}
\toprule
Key &  Average Similarity $\pm$ SV \\
\midrule
Domains & 0.266300 $\pm$  0.250173 \\
User-agents & 0.075917 $\pm$  0.240749 \\
Hostname & 0.175000 $\pm$  0.352959 \\
TLS Issuers & 0.531568 $\pm$  0.382723 \\
\bottomrule
\end{tabular}

\label{tab:similarity}
\end{table}

 \section{Related Work}
\label{sec:related}

The field of device identification is vast and diverse, and various strategies and techniques have been developed to address different aspects of this problem. The majority of  work focuses on identifying known or previously seen devices. Also, some previous researches focus on the identification of specific actions (turn on/off, etc.) \cite{sivanathan2020managing} or a specific unique model of a device, all presented in Table \ref{tab:my-table}. 

In contrast, this paper focuses on the task of labeling the vendor and the function of unknown or previously unseen IoT devices.  Existing methodologies for device identification rely on labeled data, meaning that prior knowledge about the devices under scrutiny is a prerequisite. These details are predominantly gathered through passive or active\cite{yang2019towards,bratus2008active,feng2018acquisitional} methods, where network samples produced by a device are monitored and subsequently used to generate an identification pattern \cite{bartlett2007understanding}. This approach has been adopted in numerous studies, as highlighted by \cite{guo2018ip,ortiz2019devicemien,hafeez2020iot,marchal2019audi,aneja2018iot,bao2020iot}.
All mentioned works identify only seen devices. Our research diverges from these conventional methodologies and addresses a more intricate and advanced issue, the labeling of previously unseen devices. 

There are a limited number of studies that have ventured into the realm of identifying unseen devices. Wang et al. \cite{wang2022wysiwyg} present a method to identify unseen devices using active probing methods. Le et al. \cite{le2019policy} proposed two algorithms for vendor identification of unseen devices; they used a blend of domain owners, certificate owners, and the OUI among other parameters, to reach a correct vendor classification rate of 72 devices out of 94 devices (76\%). 

Type classification was done on seen data only with an extensive training phase, which is both costly and labor intensive. %\anat{ BAR --- IT IS NOT CLEAR WHY WE ARE BETTER THAN THEM!!!! YOU TOLD ME THAT THEY ONLY DO VENDOR AND NOT FUNCTION !!! YOU NEED TO EXPLAIN WHY WE ARE BETTER } 
Bai et al. \cite{bai2018automatic} uses ML, specifically time-series, to classify device types. They segmented each device traffic into fixed-time sub-flows. The final dataset consisted of 15 devices belonging to 4 categories and achieved 75\% accuracy.
%\anat{ BAR -- AGAIN... HOW FROM TIME SERIES YOU CAN KNOW LABELS ??????????????? MAYBE THEY DIDN'T DO VENDORS ????}
Table \ref{tab:my-table}, presents a summary of several papers, organized into the 3 categories (data, output, and methodology).

We note that in the IETF's Manufacturer Usage Description (MUD) framework \cite{mud_ietf}, the IoT device self-identifies, and an allow-list (access control list) is fetched. However, there has been no real-world adoption of MUD to date.

\begin{table}[ht]
\centering
\begin{tabular}{|l|l|l|l|}
\hline

  & \textbf{Data} & \textbf {Output} & \textbf{Methodology}
  \\
 
\textbf{Paper} & {\begin{tabular}[c]{@{}l@{}}Seen(S)\textbackslash\\ Unseen(U) \end{tabular}} & {\begin{tabular}[c]{@{}l@{}}Type(T) \textbackslash\\  Vendor(V) \textbackslash \\ Unique (U) \textbackslash{} \\ Actions (A)\end{tabular}} & {\begin{tabular}[c]{@{}l@{}}Active (A) \textbackslash\\ Passive(P)\end{tabular}} \\ 
\hline
\textbf{\cite{le2019policy}} & \begin{tabular}[c]{@{}l@{}}S\\ U\end{tabular} & \begin{tabular}[c]{@{}l@{}}T\\ V \end{tabular} & \begin{tabular}[c]{@{}l@{}}P\\ P \end{tabular} \\ \hline
\textbf{\cite{bai2018automatic}} & U & T & P \\ \hline
\textbf{\cite{ammar2019network}} & S & T & P+A \\ \hline
\textbf{\cite{bao2020iot}} & S & T & P \\ \hline
\textbf{\cite{miettinen2017iot}} & S & U & P \\ \hline
\textbf{\cite{msadek2019iot}} & S & U & P \\ \hline
\textbf{\cite{wang2022wysiwyg}} & U & T+V & A \\ \hline
\textbf{\cite{marchal2019audi}} & S & U & P \\ \hline
\textbf{\cite{ortiz2019devicemien}} & S & U & P \\ \hline
\textbf{\cite{sivanathan2019inferring}} & S & U & P \\ \hline
\textbf{\begin{tabular}[c]{@{}l@{}}\cite{sivanathan2020managing}\end{tabular}} & S & U+A & P \\ \hline
\textbf{\cite{saidi2020haystack}} & S & U & P \\ \hline
\textbf{\cite{khandait2021iothunter}} & S & T+V & P \\ \hline
\textbf{\cite{hamza2021clustering}} & S & U & P \\ \hline
\textbf{\cite{yang2019towards}} & S & T+V+U & A \\ \hline
\end{tabular}
\caption{Existing IoT identification mechanisms based on 3 key categories data (Seen vs Unseen), output (Type, Vendor, Unique, specific Actions) and methodology (Active vs Passive).}
\label{tab:my-table}
\end{table}
 
\section{Conclusion \& Future Work}

This study demonstrates that, by combining recent advances in LLM and NLP, we can effectively label unseen IoT devices in the wild. Over a set of 97 unique IoT devices, our function labeling approach achieved HIT1 and HIT2 scores of 0.7 and 0.81, respectively. 

In the IoT realm, it is generally not feasible to obtain samples from every type. We used zero-shot classification models, leveraging the fact that it allows the update of the labels with no need of re-training the model itself. As far as we know, this work is the first to address the IoT device labeling problem. 

For future work, there is room for exploration in the realm of LLM explainability. We propose additional refinements to the LLM to deliver improved explainability by fine-tuning the model over multiple examples with human-written explanations. The expected outcome is that these advancements will not only facilitate a deeper understanding of the labeling procedure but will also foster greater confidence in the output of the model.

We also plan to ascertain the validity of the explanations provided by our model. By manually tagging these explanations, we can measure their semantic similarity to human-provided explanations by employing a regression model such as DistilBERT~\cite{sanh2020distilbert} as the scoring function~\cite{chen2023frugalgpt}. Quantifying this similarity will allow us to assign a score to each explanation, providing a reference for future improvements to our model. 

We also intend to explore refinements in the features that are collected and enriched.  For example, some of the domain names are AWS services that are used by many vendors, and they distract the labeling algorithm. In our ongoing work, we identify these domains and omit them from the catalog of features.
 % ANAT
 
\bibliographystyle{plain}
%\bibliography{\jobname}
\bibliography{main}

%\appendix

%\input{prompt}
\end{document}